\documentclass[apj, numberedappendix]{emulateapj}

\usepackage{xspace}
\usepackage[usenames,dvipsnames]{xcolor}
\usepackage{hyperref}
\usepackage{afterpage}
\usepackage{multirow}
\usepackage{threeparttable}

\hypersetup{pdftitle= {A Direct Measurement of Bias and Dispersion for Type Ia Supernovae Lightcurve Fitting: Limitations of Implicit K-corrections}, 
pdfauthor = {Clare Saunders},
colorlinks=true,        
linkcolor=blue,         
citecolor=blue,         
urlcolor=blue           
 }

\begin{document}
\title{Type Ia Supernova Distance Modulus Bias and Dispersion from K-correction errors: \\A direct measurement using Lightcurve Fits to Observed Spectral Time Series }

\shorttitle{Bias and Dispersion in SN Ia Lightcurve Fits}
\author{
    C.~Saunders,\altaffilmark{1,2}
    G.~Aldering,\altaffilmark{1}
    P.~Antilogus,\altaffilmark{3}
    C.~Aragon,\altaffilmark{1}
    S.~Bailey,\altaffilmark{1}
    C.~Baltay,\altaffilmark{4}
    S.~Bongard,\altaffilmark{3}
    C.~Buton,\altaffilmark{5}
    A.~Canto,\altaffilmark{3}
    F.~Cellier-Holzem,\altaffilmark{3}
    M.~Childress,\altaffilmark{1,2}
    N.~Chotard,\altaffilmark{5}
    Y.~Copin,\altaffilmark{5}
    H.~K.~Fakhouri,\altaffilmark{1,2}
    U.~Feindt,\altaffilmark{6}
    E.~Gangler,\altaffilmark{5}
    J.~Guy,\altaffilmark{3} 
    M.~Kerschhaggl,\altaffilmark{6}
    A.~G.~Kim,\altaffilmark{1}
    M.~Kowalski,\altaffilmark{6}
    J.~Nordin,\altaffilmark{1}
    P.~Nugent,\altaffilmark{7,8}
    K.~Paech,\altaffilmark{6}
    R.~Pain,\altaffilmark{3}
    E.~Pecontal,\altaffilmark{9}
    R.~Pereira,\altaffilmark{5}
    S.~Perlmutter,\altaffilmark{1,2}
    D.~Rabinowitz,\altaffilmark{4}
    M.~Rigault,\altaffilmark{6}  
    D.~Rubin,\altaffilmark{1,2}
    K.~Runge,\altaffilmark{1}
    R.~Scalzo,\altaffilmark{10}
    G.~Smadja,\altaffilmark{5}
    C.~Tao,\altaffilmark{11,12}
    R.~C.~Thomas,\altaffilmark{7}
    B.~A.~Weaver,\altaffilmark{13}
    C.~Wu\altaffilmark{3,14}
    (The Nearby Supernova Factory)}
\altaffiltext{1}
{    Physics Division, Lawrence Berkeley National Laboratory, 
    1 Cyclotron Road, Berkeley, CA, 94720}
\altaffiltext{2}
{    Department of Physics, University of California Berkeley,
    366 LeConte Hall MC 7300, Berkeley, CA, 94720-7300}
\altaffiltext{3}
{    Laboratoire de Physique Nucl\'eaire et des Hautes \'Energies,
    Universit\'e Pierre et Marie Curie Paris 6, Universit\'e Paris Diderot Paris 7, CNRS-IN2P3, 
    4 place Jussieu, 75252 Paris Cedex 05, France}
\altaffiltext{4}
{    Department of Physics, Yale University, 
    New Haven, CT, 06250-8121}
\altaffiltext{5}
{    Universit\'e de Lyon, F-69622, Lyon, France ; Universit\'e de Lyon 1, Villeurbanne ; 
    CNRS/IN2P3, Institut de Physique Nucl\'eaire de Lyon.}
\altaffiltext{6}
{    Physikalisches Institut, Universit\"at Bonn,
    Nu\ss allee 12, 53115 Bonn, Germany}
\altaffiltext{7}
{    Computational Cosmology Center, Computational Research Division, Lawrence Berkeley National Laboratory, 
    1 Cyclotron Road MS 50B-4206, Berkeley, CA, 94720}
\altaffiltext{8}
{    Department of Astronomy, University of California Berkeley, 
     B-20 Hearst Field Annex \# 3411, Berkeley, CA, 94720-3411 }
\altaffiltext{9}
{    Centre de Recherche Astronomique de Lyon, Universit\'e Lyon 1,
    9 Avenue Charles Andr\'e, 69561 Saint Genis Laval Cedex, France}
\altaffiltext{10}
{    Research School of Astronomy and Astrophysics,
    The Australian National University,
    Mount Stromlo Observatory,
    Cotter Road, Weston Creek ACT 2611 Australia}
\altaffiltext{11}
{    Centre de Physique des Particules de Marseille,
 Aix-Marseille Universit\'e,
  CNRS/IN2P3, 163, Avenue de Luminy - Case 902 - 13288 
Marseille Cedex 09, France}
\altaffiltext{12}
{    Tsinghua Center for Astrophysics, Tsinghua University, Beijing 100084, China }
\altaffiltext{13}
{    Center for Cosmology and Particle Physics,
    New York University,
    4 Washington Place, New York, NY 10003, USA}
\altaffiltext{14}
{    National Astronomical Observatories, Chinese Academy of Sciences, Beijing 100012, China}
\shortauthors{Saunders et al.}

\begin{abstract}
We estimate systematic errors due to K-corrections in standard photometric analyses of high redshift Type~Ia supernovae. Errors due to K-correction occur when the spectral template model underlying the lightcurve fitter poorly represents the actual supernova spectral energy distribution, meaning that the distance modulus cannot be recovered accurately. In order to quantify this effect, synthetic photometry is performed on artificially redshifted spectrophotometric data from 119 low-redshift supernovae from the Nearby Supernova Factory, and the resulting lightcurves are fit with a conventional lightcurve fitter. We measure the variation in the standardized magnitude that would be fit for a given supernova if located at a range of redshifts and observed with various filter sets corresponding to current and future supernova surveys. We find significant variation in the measurements of the same supernovae placed at different redshifts regardless of filters used, which causes dispersion greater than $\sim0.05$~mag for measurements of photometry using the Sloan-like filters and a bias that corresponds to a $0.03$ shift in $w$ when applied to an outside data set. To test the result of a shift in supernova population or environment at higher redshifts, we repeat our calculations with the addition of a reweighting of the supernovae as a function of redshift and find that this strongly affects the results and would have repercussions for cosmology. We discuss possible methods to reduce the contribution of the K-correction bias and uncertainty.
\end{abstract}
\keywords{Cosmology: observations --- Supernovae: general}

\section{Introduction}
\label{sec: Intro}
For several years now, Type~Ia supernovae have no longer been limited by statistical uncertainty in their use as standardized candles. Therefore, in order to improve their use as cosmological tools, it has become necessary to further limit systematic errors to obtain the highest possible accuracy from supernova measurements. This becomes increasingly crucial as we find more supernovae to constrain the properties of dark energy. Modern supernova searches (such as the Supernova Legacy Survey, \citealt{Astier:2006}, and the Dark Energy Survey, \citealt{Bernstein:2012}) typically are designed to find supernovae in a certain redshift range using a fixed set of photometric filters, which means that the supernovae found will be observed with a range of supernova-frame filter configurations. To use these supernovae for cosmology, the observations generally are converted into peak standardized absolute magnitudes in a common band.

This conversion necessarily involves using an estimate of the spectral time series. Originally, photometric observations were converted to a common band by adding an explicit K-correction to the magnitudes, which was calculated with an example spectrum (a process originally developed for use on galaxies in \cite{Humason:1956} and \cite{Oke:1968}). \cite{Kim:1996} extended this method with the development of the cross-filter K-correction, which allowed for conversion between different filters. K-corrections were further improved by \cite{Nugent:2002fk} with a spectral time series template that could be stretched or warped to match a supernova's shape. Once the supernova observations have been K-corrected, they can be fit by a lightcurve template to get the standardized peak absolute magnitude, as done by MLCS2k2 or SNooPy (\citealt{Jha:2007ys}, \citealt{Burns:2010vn}).

Modern spectral time series-based lightcurve fitters combine the two-step process of K-correcting and then fitting photometric data. Lightcurve fitters like SALT2 and SiFTO (\citealt{Guy:2007fk}, \citealt{Conley:2008kx}) fit observed photometric data with integrals over a time-evolving spectral energy distribution model that can be adjusted by means of free parameters. The standardized magnitude is calculated from the parameters and the rest-frame magnitude of the best-fit model.

In either of these methods, errors in the supernova's magnitude will occur if the estimated spectrum is different from the true spectrum of the observed supernova, either in the spectral shape or the flux normalization with respect to the other phases. This can happen if the diversity of the supernova population is greater than can be encompassed by the model. Further, since the estimated spectrum is based on the available photometry, it can vary as a function of the filters used to observe the supernova, depending on which spectral features fall in each filter band and the relative weighting of different parts of the spectral time series model. While a separate K-correction is only explicitly done in the first type of lightcurve fitter, we will here refer more generally to errors due to inaccurate estimation of the supernova spectrum as K-correction errors.

Variation in the best-fit spectrum propagates into variation in the distance modulus. Uncertainty due to a random scatter in the errors will decrease in proportion to the square root of the number of supernovae observed. However, bias in the ensemble mean of the magnitudes will cause a systematic error in the fitting of the Hubble Diagram that will not be reduced with greater numbers of supernovae. Such errors are thought to be subdominant, but have not been exactly tested using a current-generation lightcurve fitter and a large supernova data set.

With photometric data alone, it would be extremely difficult to cleanly measure the size of this effect, but using supernova spectrophotometric time series data it is possible to exactly measure the effect on each supernova. Spectrophotometry can be integrated over any passband to synthesize photometry for the same supernova as if it were observed through different filters or at different redshifts; changes in the lightcurve fit for different filters due to K-correction error are then directly measurable. From the ensemble of supernovae in the data set, the bias and dispersion on that supernova population can be calculated. In this paper, we calculate the redshift-dependent distance modulus variation due to K-corrections by applying a state-of-the-art two-parameter lightcurve fitter to spectrophotometric time series data from the Nearby Supernova Factory (SNfactory, \citealt{Aldering}). The bias and distribution of distance modulus variation over the whole ensemble provide an estimate of the added uncertainty in distance modulus applicable to other supernovae without spectroscopic time series that will be used in cosmological measurements.

In Section~\ref{sec: data}, the SNfactory supernova data set is described. The questions we address with different filter sets are given in Section~\ref{sec: Filters}, with a description of the procedure used in Section~\ref{sec: Proc}. Section~\ref{sec: Sources} compares the K-correction effects when using these different sets of photometric filters and discusses the effect of potential hidden calibration errors in the sample supernova set. In Section~\ref{sec: contrib} we consider the contribution of K-correction effects to the total dispersion seen in supernova standardized magnitudes. We also consider the applications of our results for future high-redshift programs, where different filter configurations are planned, and with the addition of possible shifts in the makeup of the supernova population in Section~\ref{sec: Apps}. Finally, in Section~\ref{sec: disc}, we discuss the sources of the errors that are seen and we discuss possible methods for minimizing errors due to K-correction.

\section{The Supernova data set} 
\label{sec: data}
The data set used in this study consists of spectrophotometric time series for a set of over 100 nearby supernovae, observed by the SNfactory between 2005 and 2009 with the SuperNova Integral Field Spectrograph (SNIFS, \citealt{Lantz:2004}). Most of these supernovae have been previously presented, for example in \cite{Thomas:2007, Thomas:2011}, \cite{Bailey:2009}, and \cite{Chotard:2011}. SNIFS is a fully integrated instrument optimized for automated observation of point sources on a structured background over the full ground-based optical window at moderate spectral resolution ($R \sim500$). It consists of a high-throughput wide-band pure-lenslet integral field spectrograph (IFS, ``\`a la TIGER;" \citealt{Bacon:1995, Bacon:2000, Bacon:2001}), a multi-filter photometric channel to image the field in the vicinity of the IFS for atmospheric transmission monitoring simultaneous with spectroscopy, and an acquisition/guiding channel. The IFS possesses a fully-filled $6.\!''4 \times 6.\!''4$ spectroscopic field of view subdivided into a grid of $15 \times 15$ spatial elements, a dual-channel spectrograph covering $3200-5200$ \AA\ and $5100-10000$ \AA\ simultaneously, and an internal calibration unit (continuum and arc lamps). SNIFS is continuously mounted on the South bent Cassegrain port of the University of Hawaii $2.2$ m telescope on Mauna Kea and is operated remotely. A description of host-galaxy subtraction is given in \cite{Bongard:2011}. 

The spectra are flux-calibrated following the procedure detailed in \cite{Buton:2012} and the color calibration is trusted to within $1\%$ based on observations of standard stars. Comparison between spectra from the same supernovae observed by SNIFS and by the Hubble Space Telescope (\citealt{Maguire:2012}) confirms that any $B-V$ color offset is below this $1\%$ level. In the analysis we have included a dispersion empirically measured in the lightcurve fits using SNfactory and other data sets as an error floor of $0.04$~mag on the ability of the lightcurve model to match the photometric data points. The SNfactory supernova spectra are originally calibrated using a hybrid system of both CALSPEC ({\citealt{Bohlin:2014} and references within) and Hamuy (\citealt{Hamuy:1992, Hamuy:1994}) standard stars, which are listed in \cite{Buton:2012}. However, in this analysis the supernovae have been recalibrated to the CALSPEC standards in order to match the calibration of the SALT2 training data. The effect of the calibration on the analysis is discussed further in Section~\ref{sec: Sources}.

The supernovae in our data set range in redshift from $z\approx 0.007$ to $0.11$ and in epoch from approximately 12 days before maximum to 55 days after, with an average of $14$ spectra for each supernova. The supernovae spectra have been corrected for Milky Way dust extinction (\citealt{Schlegel:1998}, \citealt{Cardelli:1989}) and placed in a common rest-frame at $z=0$. In order to preserve any characteristics caused by circumstellar and host galaxy extinction, we do not correct for any further reddening due to these dust sources. We do not include in our sample a small number of supernovae that have been previously identified as Ia-CSM or super-Chandrasekhar mass Type Ia supernovae and that would be easily identified and removed from a photometric supernova survey (\citealt{Aldering:2006}, \citealt{Scalzo:2010}, \citealt{Scalzo:2012}). After removing these, we are left with 119 supernovae in our set, which are listed in \citealt{Childress:2013}.

\section{Questions addressed and filters used}
\label{sec: Filters}
Different sets of photometric filters, shown in Figure~\ref{fig: Filters} and described in Table~\ref{table:filters}, are used to answer a series of questions.
\begin{figure}
\includegraphics[width = \columnwidth]{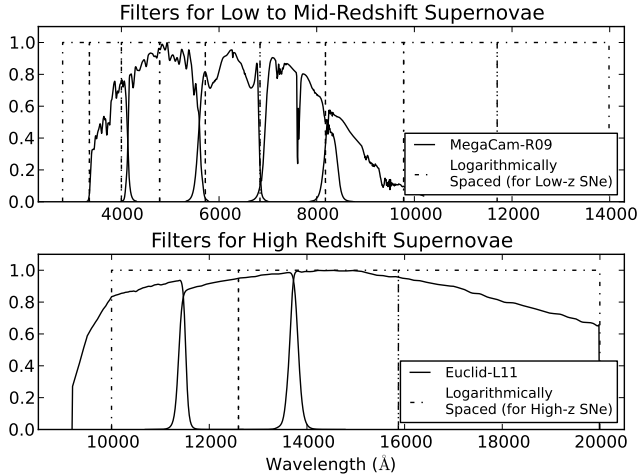}
\caption{Transmission of filters sets used for synthetic photometry. The MegaCam-R09 (Sloan-type) filters and Euclid-L11 filters are shown with the logarithmically-shaped filters covering similar wavelength ranges.}
\label{fig: Filters}
\end{figure}

\begin{table*}
\begin{threeparttable}
\caption{Filters Set Specifications}
\begin{center}
\begin{tabular}{ c c  c  c c}
\hline\hline
Filter Set & Filter Name & Wavelength Range (\AA) & Central Wavelength (\AA) & Redshift Line-up\tnote{b}\\
\hline
\multirow{8}{3.5cm}{Logarithmically-Spaced (for low-redhift SNe)}  & LL1 & 3345 -- 4000 & 3672 &\\
 & LL2 & 4000 -- 4783 &  4391 & LL1$\cdot (1+0.20)$
 \tnote{c}\\
 & LL3 & 4783 -- 5719& 5251 &  LL1$\cdot (1+0.43)$\\
 & LL4 & 5719 -- 6839 & 6279 & LL1$\cdot (1 + 0.71)$\\
 & LL5 & 6839 -- 8179 & 7509 &   LL1$\cdot (1 + 1.05)$\\
 & LL6 & 8179 -- 9780 & 8979 &  LL1$\cdot (1+1.45)$ \\
 & LL7\tnote{a} & 9780 -- 11696 & 10738 &  LL1$\cdot(1+1.92)$\\
 & LL8\tnote{a} & 11696 -- 13987 & 12841&  LL1$\cdot(1+2.50)$\\
\hline
\multirow{5}{*}{MegaCam-R09} & u & 3370 -- 4110 & 3740 & \multirow{5}{*}{None}\\
  & g & 4140 -- 5590 & 4870 \\
  & r & 5640 -- 6850 & 6250 \\
  & i & 6980 -- 8430 & 7700 \\
 & z & 8230 -- 10190 & 8900 \\
  \hline 
\multirow{3}{3.5cm}{Logarithmically-Spaced  (for high-redshift SNe)} & LH1 & 10000 -- 12600 & 11300\\
 & LH2 & 12600 -- 15873 & 14236 &  LH1$\cdot(1.26) = $ LH1$\cdot\frac{(1 + 1.52)}{(1+1.0)} $ \tnote{d}\\
 & LH3 & 15873 -- 20000 & 17936 & LH1$\cdot (1.59) = $ LH1$\cdot\frac{(1 + 2.17)}{(1+1.0)}$ \\
 \hline
\multirow{3}{*}{Euclid-L11} & Y & 9200 -- 11460 & 10330 & \multirow{3}{*}{None}\\
  & J & 11460 -- 13720 & 12590\\
  & H & 13720 -- 20000 & 16860 \\
\hline
\end{tabular}
\begin{tablenotes}
\item[a] Not included in ``finite" set of logarithmically-spaced filters.
\item[b] Determined by the redshifts at which $T_{A}(\lambda\cdot (1+z_{1})) = T_{B}(\lambda \cdot (1+z_{2}))$, where $T_{A}$ is the transmission for filter $A$ as a function of $\lambda$. 
\item[c] As an example, we show the redshifts at which the supernova-frame LL1 band lines up with each of the other bands at $z=0$. 
\item[d] Here we show the supernova-frame redshifts at which the LH1 band lines up with the other bands at the arbitrarily chosen supernova-frame redshift $z=1$.

\end{tablenotes}
\end{center}
\label{table:filters}
\end{threeparttable}
\end{table*}

In order to test the best-case scenario where observer-frame and supernova-frame filters can be exactly matched, in Section~\ref{sec: Sources} we use a set of top-hat shaped filters with logarithmically spaced edges, which extend to whatever wavelength range is needed to fully cover the supernova spectrum and so are called the ``infinite" set (top panel of Figure~\ref{fig: Filters} and Table~\ref{table:filters}). The filter spacing is chosen so that the LL4, LL5, and LL6 bands approximately match bands $r$, $i$, and $z$ in the MegaCam-R09 filter set, allowing for a closer comparison between them. Because of the logarithmic spacing, these filters move in and out of alignment across redshift such that, at certain redshifts, filters in the observer frame exactly overlap with filters in the supernova frame. A similar technique was used in \cite{Schmidt:1998}, where filters meant to mimic $B$ and $V$ filters redshifted to $z = 0.35 $ and $z=0.45$ were used to minimize K-correction errors in the range $0.25 < z < 0.55$. The logarithmically-spaced filter set used here presents a more flexible variation on the redshifted $B$ and $V$ filters, since all of the filters in this set can line up with each other, extending the redshift ranges where close matches can be made between supernova-frame and observer-frame filters. Because of this characteristic, along with their on-off edges and full coverage of the wavelength range used by lightcurve fitters, the logarithmically-spaced filters are an idealized set. While narrower filters would provide more information, this would be balanced by a trade-off in increased observing time, meaning overall the filter configuration used here is approximately optimal.

From this we proceed to two degraded situations. To simulate the effect of a finite filter set losing coverage of the redder end of the supernovae's spectral energy distribution, we restrict the analysis to the logarithmically-spaced filters covering $3300$ to $9800$ \AA, referred to as the ``finite" set. Next, to simulate the effects of having filters that do not align with each other at any redshift, we use the MegaCam (Sloan-like) filters (also shown in the top panel of Figure~\ref{fig: Filters}), which were used in the Supernova Legacy Survey (hereafter referred to as MegaCam-R09, \citealt{Regnault:2009}). The MegaCam-R09 filter transmission curves are calculated at $14$~cm from the center of the focal plane and include average atmospheric transmission at Mauna Kea, the CCD quantum efficiencies, the transmission along the optical path, and the mirror reflectivity. In Section~\ref{sec: Apps}, we address the issue of fitting high-redshift supernovae, where different filters need to be used for the measurements' low-redshift anchor. For higher-redshift supernovae, we use the filters planned for the original design of the Euclid Space Telescope (referred to as Euclid-L11 filters, \citealt{Laureijs:2011}), the results of which are then contrasted with an artificial logarithmically-spaced filter set we have designed to have a similar wavelength coverage (both shown in the bottom panel of Figure~\ref{fig: Filters}).

\section{Procedure}
\label{sec: Proc}
In this analysis we use SALT2 (both version 2.2, \citealt{Guy:2010}, and the recently released version 2.4, \citealt{Betoule:2014}) as an exemplar of the capabilities of a state-of-the-art two-parameter lightcurve fitter. Other lightcurve fitters based on similarly simple templates are expected to give comparable results (SiFTO, for example, was shown in \cite{Conley:2008kx} to be very similar to SALT2). Using the spectrophotometric data for each of the SNfactory supernovae, we construct mock supernovae at fixed redshift intervals, using a given set of filter functions to perform synthetic photometry. The supernova spectra are shifted in wavelength, but the flux normalization of the supernovae is not changed because we are interested in the differences in lightcurve fits due to filter coverage, not supernova distance.

We begin by using SALT2 to fit the multi-band lightcurves of a supernova at $z=0$ and use this fit as a zero point to which other fits of the supernova are compared. SALT2 fits a spectral energy distribution model to the supernova data by using two free parameters that correspond to the color and lightcurve shape of the supernova. By combining the fitted peak B band magnitude with these color and $x_{1}$ (stretch-like) parameters, we get the standardized magnitude, $M_{std}$:
\begin{equation}
M_{std} = M_{B} + \alpha \; x_{1}- \beta \; c
\end{equation} 
where $\alpha$ and $\beta$ are redshift-independent terms that can be fit by minimizing the residuals in the Hubble diagram (in this analysis we use $\alpha=0.137$ and $\beta = 3.07$, \citealt{Rubin:2013} {to emulate cosomological analyses}). We stress that this standardization method, including the $\alpha$ and $\beta$ parameters, is entirely extrinsic to the SALT2 lightcurve fit; it is only the systematic variations on $M_{B}$, $x_{1}$ and $c$ that concern us here. We then fit the lightcurve of the redshifted version of the same supernova. As discussed in Section~\ref{sec: Intro}, in order to fit this lightcurve and get the standardized peak magnitude of the redshifted supernova, a K-correction must be involved in fitting a template to the supernova, which necessarily uses an approximation (in the form of an imperfect spectral model) to fit the measured data. This introduces error into the calculation. Having set the standardized magnitude found for the rest-frame as our zero point, we can find the fit-dependent difference between the two distance estimators of the supernova:
\begin{equation}
\Delta \mu(z_{i}) \equiv M_{std | z=z_{i}} - M_{std | z=0}
\end{equation}
This allows us to see the variation around the expected standardized magnitude (or equivalently the fit-dependent part of the variation in the distance modulus) found when using a two-parameter model to fit the same supernova placed at different redshifts. For simplicity, we call this difference the error due to K-correction.

\begin{figure}
\includegraphics[width = \columnwidth]{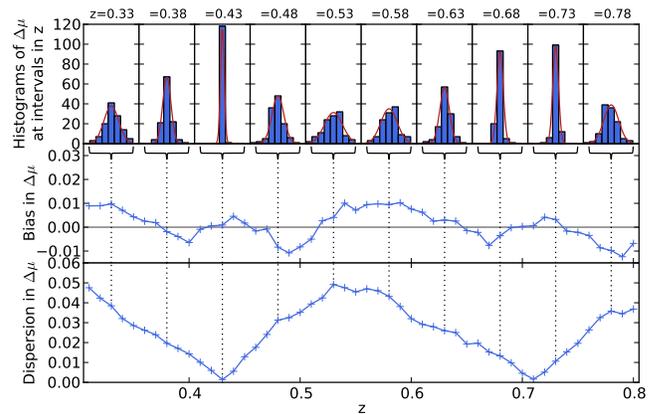}
\caption{An example of the median bias and dispersion in the standardized magnitude variation due to K-correction errors found using SALT2.2 (with logarithmically-spaced filters) over a portion of the redshift range are shown in the middle and bottom panels. The error in the bias is approximately $\sqrt{\pi/(2N)}\approx 0.11$ times the dispersion. The top panel shows histograms of the data points at intervals of $0.05$ in redshift, where each histogram has 9 bins spanning standardized magnitude errors from $-0.15$ to $0.15$. A Gaussian curve fit to the histogram is overplotted in each panel. Low standard deviation in the histogram can be seen to correspond to low dispersion in the bottom panel, while a large standard deviation corresponds to high dispersion.}
\label{fig: Hist}
\end{figure}

At a given redshift, the above procedure is repeated for the 119 supernovae in our data set, and from the aggregate of the errors we assess the accuracy of the K-corrections. Examples of the range of errors across redshifts may be seen in the top panel of Figure~\ref{fig: Hist}, which shows histograms of the errors for each of the supernovae at redshift intervals. As a baseline we use robust statistics, meaning that the bias and dispersion are calculated from the median and normalized median absolute deviation (nMAD) of the errors (i.e. the median and nMAD of the $\Delta\mu$ values for all the supernovae at a given redshift), ensuring results are not sensitive to outliers. However, the mean and standard deviation of a Gaussian function fit to the histogram of the errors at a given redshift give equivalent results, as may be seen by comparing the Gaussian curves in the top panel to the amount of dispersion in the bottom panel of Figure~\ref{fig: Hist}. A bootstrap test confirms that the error in the bias is approximately equal to $\sqrt{\pi/(2N)} \times$ the dispersion, as expected for bias calculated from the median, where $N=119$, the number of supernovae used. This further shows that outliers are not driving the dispersion seen in the results.

Some supernovae are fit consistently by the template, such that their standardized magnitudes are reconstructed well; this means that even at redshifts where there is a large standard deviation, these supernovae are still in the peak of the histogram. Less well fit supernovae have standardized magnitudes which vary more across redshifts and make up the wings of the histograms at redshifts where there is a high standard deviation. The errors assembled across redshifts allow us to gauge general trends in the standardized magnitudes, in particular the amount of dispersion at different redshifts, or consistent positive or negative biases.
\begin{figure}
\includegraphics[width = \columnwidth]{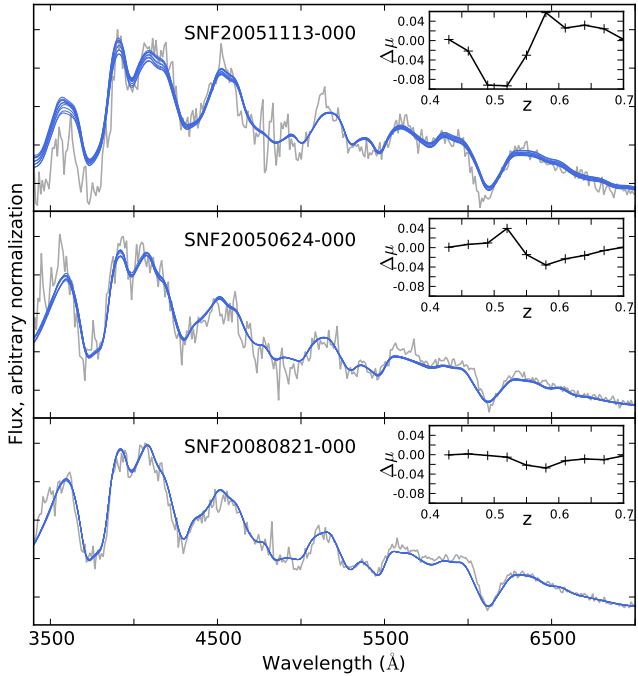}
\caption{Examples of spectral reconstructions of different qualities of SALT2.2 lightcurve fits calculated when using logarithmically-spaced filters. Spectra near maximum for three example supernovae are shown in gray. Overplotted are the spectra constructed from the SALT2 parameters fit to the supernova at a range of redshifts. The inset plots show the values for the variation in the standardized magnitudes at the corresponding redshifts. The top panel shows a supernova for which the fits vary widely, the middle panel shows a supernova for which there is moderate variation, and the bottom panel shows a supernova which SALT2 fits very similarly at any redshift.}
\label{fig: Spectra}
\end{figure}

As an example of the possible variation of SALT2 lightcurve fits, in Figure~\ref{fig: Spectra} we show the rest-frame spectra nearest maximum of three supernovae, along with the SALT2 model spectra constructed from the parameters fit to each supernova at a range of redshifts. It can be seen that for the supernova in the top panel, the best-fit SALT2 spectral reconstructions vary noticeably across redshift, while for the supernova in the bottom panel, the SALT2 reconstructions are more difficult to distinguish from one another.

\section{Sources of K-correction Errors}
\label{sec: Sources}
We consider first the results of our analysis when using version 2.2 of SALT2: For all supernovae, when an object is measured with logarithmically-spaced filters, shown as dotted lines in Figure~\ref{fig: Filters}, which fully cover the wavelength range used by SALT2, there are some redshifts where the filters in the frame of the supernova and the filters in the frame of the observer line up exactly. When a supernova is measured at one of these redshifts, with the same number of filters as were used for the rest-frame measurement, the K-correction is done in exactly the same way for the $z=0$ and the redshifted supernova, which means that the calculated standardized magnitudes are the same.

When we move away from this ideal situation, errors start to appear. The most easily identifiable source of error is when filters are out of alignment. 
\begin{figure}
\includegraphics[width = \columnwidth]{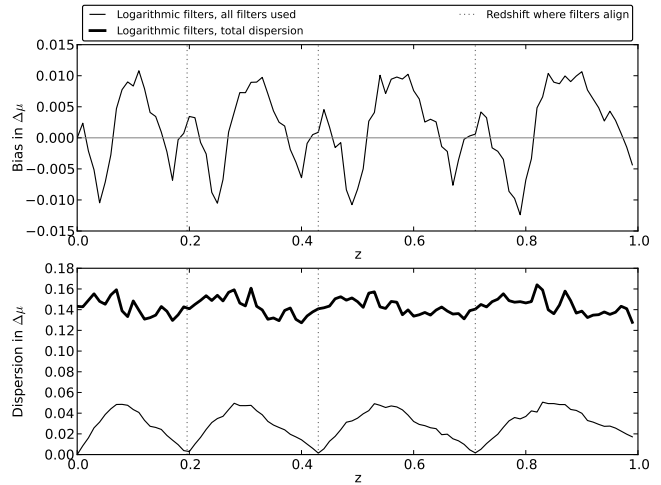}
\caption{ Bias and dispersion in the standardized magnitude variation due to K-correction errors when an ``infinite" set of logarithmic filters are used with SALT2.2. Note that these are the same results as in Figure~\ref{fig: Hist}, shown over a greater redshift range. As a comparison, the lower panel shows, the total dispersion in the standardized magnitudes, which is what would actually be observed (this is equivalent to the dispersion in the distance moduli on a Hubble diagram).}
\label{fig: stage 1}
\end{figure}
As can be seen in Figures~\ref{fig: Hist} and~\ref{fig: stage 1}, the dispersion in the errors increases and the bias oscillates as the  filters fall out of alignment, and then both go to zero as the filters become more aligned. When the filters are maximally misaligned, the dispersion peaks at about $0.04$ to $0.05$ magnitudes, while the bias varies between $-0.01$ and $0.01$ mag. In both the bias and the dispersion we see a periodically repeating pattern, the size of which does not increase as we go up to higher redshifts. It should be noted that in a cosmological analysis, a periodic bias such as this can in principle be fit out, whether or not the size of the bias is known, though in practice this has never been done.

For the subset of supernovae that are well fit by the SALT2 templates, the SALT2 fit is almost the same at every redshift (the supernova in the bottom panel of Figure~\ref{fig: Spectra} is an example of such a supernova). This source of error is thus the result of supernova spectral diversity not captured by the lightcurve fitter parametrization of the underlying spectrum. The dispersion is set by the degree of diversity, while the bias reflects a shift between the demographics of the SNfactory data set and that of the supernova set used to train the SALT2.2 spectral templates.

A further source of error occurs if we no longer have enough filters to always cover the wavelength range of the supernova at a given redshift. This means that as we get to higher redshifts, we run out of filters and measure a supernova with fewer filters than we had for the rest-frame measurement. To test this situation, the filters used are limited to those covering observer-frame 3345 to 9780 \AA, the ``finite" set. We then calculate the standardized magnitudes that would be found with the limited filters at a given redshift, compare them to the rest-frame magnitudes, and find the bias and dispersion in the aggregate of the errors (Figure~\ref{fig: stage 3}). Filter coverage of the redder wavelengths is lost starting at $z=0.53$ since the LL7 filter which would otherwise cover this range is not included in the ``finite" set. As the lack of overlapping wavelength coverage becomes more severe around $z=0.7$, a large amount of dispersion appears, increasing to more than $0.15$ magnitudes, and the bias oscillates over a much larger range. The results suggest that some loss of filter coverage can be tolerated, but that it will become a major source of systematic errors when the filter coverage is limited and covers bluer supernova-frame wavelengths. The possible causes of this effect are discussed below.

A starker difference in the standardized magnitudes is seen if filters are used which are not logarithmic in their shape and relative spacing, as is shown in Figure~\ref{fig: stage 3}. 
\begin{figure}
\includegraphics[width = \columnwidth]{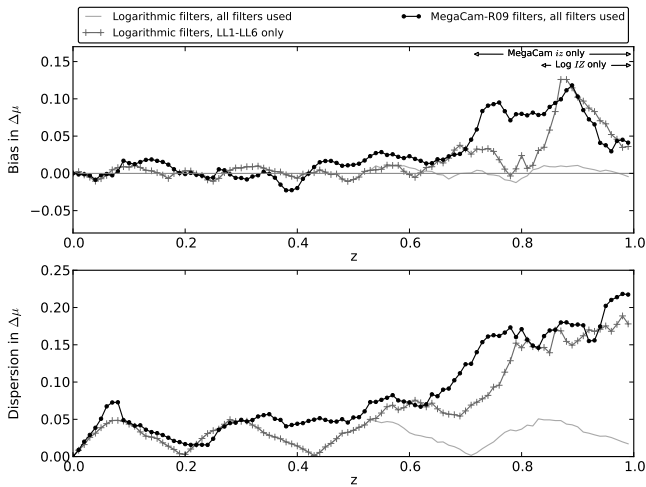} 
\caption{Bias and dispersion in standardized magnitude variation found using the MegaCam-R09 filters with SALT2.2. Results from previous iterations using an ``infinite" and a finite number of logarithmically-shaped filters are shown in light and dark gray. The redshift ranges in which only two MegaCam-R09 or Logarithmic filters are available with the SNfactory data are indicated in the top panel.}
\label{fig: stage 3}
\end{figure}
As a comparison to the logarithmically-spaced filter set, we look at the filters used in the MegaCam wide-field imager, a set of Sloan-like filters which are shown in Figure~\ref{fig: Filters} and have wavelength coverage similar to that of the ``finite" set of logarithmic filters. Here there is no longer any redshift where the supernova-frame filters line up with the observer-frame filters and thus there are no redshifts at which a K-correction can be done perfectly. Instead, there is a consistent amount of dispersion at around $0.05$ magnitudes, increasing to about $0.20$~mag as wavelength coverage is lost. We find a small bias in the data, varying across redshift within about $0.02$~mag between $z=0$ and $0.7$, but by an increasing amount as redder wavelength coverage is lost at higher redshifts.

We next consider the results of our analysis when using the recently released version 2.4 of SALT2 on synthetic photometry from the same three filter sets. SALT2.4 uses the same model as SALT2.2, but the templates and color law have been retrained on a larger data set, which changes both the individual fits and the ensemble average, and the error bands have been improved. As can be seen in Figure~\ref{fig: SALT2-4}, at redshifts below $z=0.6$, the results are very similar to those found with version 2.2. However, at higher redshifts, the bias is much lower than was found when using 2.2. The dispersion, on the other hand, is very similar to that found using SALT2.2. Reasons for this are discussed below.

\begin{figure}
\includegraphics[width = \columnwidth]{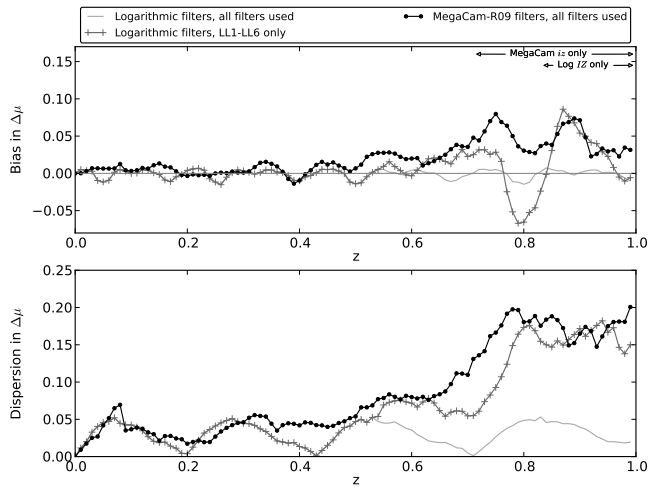} 
\caption{Same as Figure~\ref{fig: stage 3} but using SALT2.4 for the lightcurve fits.}
\label{fig: SALT2-4}
\end{figure}

\subsubsection*{Isolating the Cause of the Results at Higher Redshifts}
The high bias and dispersion seen above $z\approx0.7$ in Figure~\ref{fig: stage 3} raise questions about lightcurve fitting when coverage of the red side of supernova's spectra is lost at higher redshifts. Here we examine possible causes for this breakdown in K-correction accuracy, such as an overall offset between the calibration of SALT2 and the SNfactory data, which would lead to biased results, or dispersion in the calibration, which would lead to dispersion in the results. 

First, there is uncertainty in the calibration of the data used to train SALT2. We perform a Monte Carlo simulation to test the impact of SALT2 input calibration errors following the error distribution shown in Appendix A of \citealt{Guy:2010}. These errors are used to produce variations in the SALT2 spectral template, which are then propagated through the K-corrections analysis by fitting the template variations with SALT2. This gives an error on the bias equal to about $0.01$ at low redshifts, increasing to about $0.04$ at $z=1$. This uncertainty is shown for both SALT2.2 and 2.4 in gray in Figure~\ref{fig: SALT2_MC}. A similar effect is shown in \cite{Mosher:2014}, where varying the SALT2 training data and scatter model is shown to result in biased Hubble diagrams.

\begin{figure}
\includegraphics[width = \columnwidth]{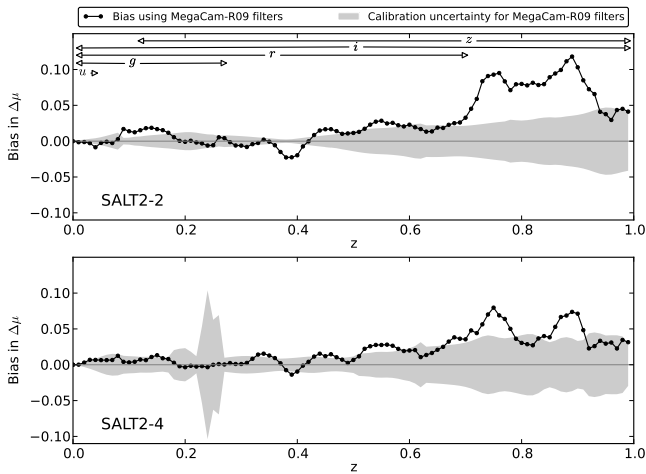}
\caption{Calibration uncertainty is shown as an error band on the bias zeropoint, overlaid on the bias calculated using SNfactory spectra. The top panel shows the bias and calibration uncertainty calculated using SALT2.2 and the MegaCam-R09 filters, while the bottom panel shows the same results calculated using SALT2.4. The spike in the uncertainty for SALT2.4 at around $z=0.25$ reflects the fact that at this redshift the $g$ band covers the far UV in the supernova-frame where there is very large calibration uncertainty in the training data. The redshift range where each of the MegaCam-R09 filters can be used is shown in the top panel; the redshift ranges for the results in the bottom panel are the same.}
\label{fig: SALT2_MC}
\end{figure}

Next, we evaluate whether undiagnosed calibration errors might play a role in the effects seen. Dispersion in the calibration of the U band (averaging to zero) would be a plausible cause of the dispersion in the standardized magnitudes, since the lightcurve fit becomes more sensitive to the U band photometry when redder bands are lost. To test whether a calibration error in the SNfactory data could cause this effect, a Monte Carlo simulation was performed in which a random, Gaussian distributed calibration error with $\sigma = 0.06$ mags was added to the U band of fake supernovae made using the SALT2 template. These supernovae were then placed at a range of redshifts and the above analysis was repeated to synthesize photometry and fit the standardized magnitude at each redshift. While SALT2 does not find exactly the same standardized magnitude at every redshift for each of these simulated supernovae, the variation is negligible in comparison to the results seen in Figure~\ref{fig: stage 3}, with dispersion at about $0.005$ mags. At higher redshifts, as redder filters are lost and the U band becomes more important when the MegaCam or finite logarithmic filters are used, the dispersion does increase to about $0.04$ mags at $z=1.0$, but this is still only a small fraction of the dispersion that we see from variation due to K-corrections (about $0.2$ mags).

This suggests that more deep-seated differences between the U bands of the SNfactory data and the SALT2 template must be the source of the large biases seen. The decrease in the biases seen between SALT2.2 and SALT2.4 in particular shows that the additional training material in the new version has brought the blue end of the SALT2 model closer to that of the average SNfactory supernova spectrum. Even with the new model the dispersion at higher redshifts has stayed high however. This implies that the UV spectra of Type~Ia supernovae are inherently more diverse than can be accounted for by a two-parameter lightcurve fitter, or that there are UV subpopulations that are not represented by the spectral templates (Figure~\ref{fig: Spectra} shows examples in which the SALT2 reconstructions do not match the spectra in the UV). Hints of the importance of subpopulations in the UV spectra have been seen already in the literature, but this will require further study (\citealt{Jha:2007}, \citealt{Ellis:2008}, Nordin et al., in preparation).

An additional calibration test was done in which a grey calibration error was added to the SALT2 template. In this scenario, a random, Gaussian-distributed wavelength-independent error with $\sigma=0.05$~mag was added to the spectrum. As in the previous test, SALT2 did not find exactly the same standardized magnitude at every redshift for the simulated supernovae. However, the bias and dispersion in these standardized magnitudes is again small compared to the bias and dispersion that we see in the fits to the real supernova data. Even with this conservative estimate of a possible grey calibration error, the dispersion is at most about $20\%$ of what is seen with the real data, and the bias is negligible until filters are lost at high redshifts, where the bias increases to about $0.004$, again small compared to the bias in the real data (compare to bias in top panel of Figure~\ref{fig: stage 3}).

The U band and grey calibration issues have been ruled out as causes of the dispersion seen at higher redshifts. Systematic errors in the standard star calibration of both the SALT2 training data and the SNfactory data cause uncertainty which may explain some of the bias that is seen. The remaining bias and the dispersion must then be due to either spectral feature differences or broad band differences in the SNfactory data set and the SALT2 training set. Visual inspection suggests that both of these are in play, but further work will be required to decide quantitatively which has the greater effect.

\section{Contribution of K-corrections to the Total Dispersion in Magnitude Measurements}
\label{sec: contrib}
The effects of K-corrections have not been seen in the data from past photometric supernova surveys because their effects are lost in what we actually observe, which is the total dispersion in supernova standardized magnitudes, or equivalently the dispersion in Hubble diagram residuals. As a comparison, the dispersion due to K-correction errors is shown with the total dispersion in the supernova standardized magnitudes in Figure~\ref{fig: stage 1}. For both logarithmically and non-logarithmically-spaced filters, the dispersion in the standardized magnitudes (as opposed to the dispersion only in the K-correction error) is close to a constant value of about $0.15$ magnitudes across redshifts, until filter coverage is lost at higher redshifts. (This is comparable to the RMS of the Hubble residuals, $0.161$ mag, found in SALT2, \citealt{Guy:2007fk}.) Thus, while K-corrections have the effect of moving individual magnitudes away from their standardized magnitude as measured in the rest-frame, these changes are subdominant to the overall spread in the data.

To test whether the results using synthetic photometry of SNfactory data agree with recent supernova searches, the total dispersion in standardized magnitudes of the redshifted SNfactory data set used here is compared to the results of the Sloan Digital Sky Survey (SDSS-II, \citealt{Holtzman}, \citealt{Frieman}, Rubin et al, in preparation), shown in Figure~\ref{fig: SDSS}. 
\begin{figure}
\includegraphics[width = \columnwidth]{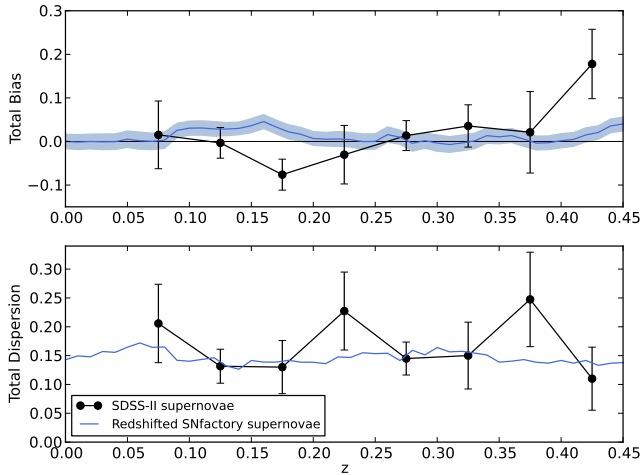}
\caption{Bias and dispersion in the Hubble residuals of supernovae from SDSS-II are shown by black points with error bars, binned at intervals of $0.05$ in redshift. The bias and dispersion in the standardized magnitudes of our mock supernova observations (where the median of the standardized magnitudes at $z=0$ is used as the zero point for the bias) are shown as a comparison by the blue line and error band. Both data sets are fit using SALT2.2.}
\label{fig: SDSS}
\end{figure}
The SDSS-II supernova set consists of 129 photometrically measured supernovae spread across redshifts $0.06$ to $0.42$. To compare the SDSS-II results with our own, the Hubble residuals of the standardized magnitudes fit by SALT2.2 (calculated with $\Omega_{m}=0.27$) were binned and then the bias and dispersion of the Hubble residuals in each bin were calculated. The variation of the bias in the SDSS-II data is on a comparable scale with the bias variation predicted from the synthetic SNfactory distance moduli. While the dispersion in this data set is not as constant, the uncertainties on the value in each bin are consistent with a constant dispersion.

Again, the amount of dispersion in the data means that the effect of K-corrections by themselves cannot be seen in the total dispersion of the data set, where they are subdominant to measurement uncertainty and intrinsic dispersion. This does not, however, mean that it is not important to isolate the K-correction errors: future advances may reduce sources of dispersion that are currently dominating what is observed. Additionally, future supernova searches will use hundreds or thousands of supernovae to narrow the statistical dispersion of binned distance moduli (\citealt{Bernstein:2012}, \citealt{LSST:2009}). Thus, while K-corrections are currently subdominant, due to the fact that they are a systematic source of bias they will become a limiting factor in precision if future searches do not address these issues, as discussed below.

\section{Implications for Supernova Cosmology}
\label{sec: Apps}
An important application of this study will be for high redshift supernova analyses, requiring that our previous results be extended to higher redshifts. Doing so for an internally-consistent set of logarithmically-spaced filters would simply extend to higher redshifts the oscillatory behavior seen in Figure~\ref{fig: stage 1}. A more interesting and relevant case is to address the effect of the jump from ground-based to space-based experiments when going to higher redshift, and the attendant filter choices. In this case, high-redshift supernovae observations will need to be compared to low-redshift supernovae that have been observed under different conditions. To calculate the K-correction errors in this situation, for the $z=0$ zero-point measurements we will use the MegaCam-R09 filters, which are a representative of ground-based experiments like SNLS, DES and LSST. For the high-redshift measurements, we use two alternative filter sets that would be realistic for a space-based mission: first, a set of three filters once planned for the Euclid mission, and second, three logarithmically-spaced filters that Euclid or other space missions could use instead (bottom panel of Figure~\ref{fig: Filters}).

Such externally-imposed changes in filter spacing mean that the ground-based and space-based rest-frame wavelength coverages never line up exactly.
\begin{figure}
\includegraphics[width = \columnwidth]{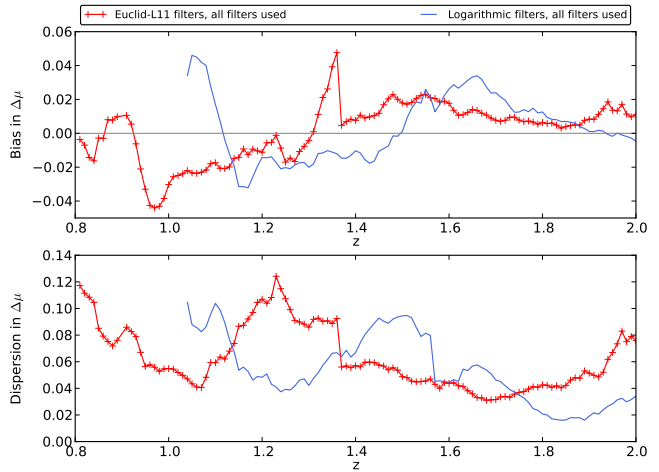}
\caption{Bias and dispersion of the standardized magnitude variation found using SALT2.2 with the Euclid-L11 filters and a set of logarithmically-shaped filters covering a corresponding wavelength range. Since these filters cannot measure low-redshift supernovae, the results from the MegaCam-R09 filters are used as the zero redshift fiducial points. The Euclid-L11 filters approximately align with the MegaCam-R09 filters in the regions around $z=1.05$ and $1.73$ and the logarithmically-spaced filters approximately align with the MegaCam-R09 filters around $z=1.27$ and $1.9$. The importance of wavelength coverage can be seen in the Euclid-L11 results, where the inclusion of the $H$ band at redshifts above $1.37$ causes a large drop in the bias and dispersion.}
\label{fig: EUC_two}
\end{figure}
This means that the K-corrections cannot be done perfectly at any redshift, but we do still see that as filters move closer and farther from approximate alignment the dispersion rises and falls (Figure~\ref{fig: EUC_two}). Though neither filter set appears particularly favored in the comparison with a MegaCam-R09 baseline, the logarithmically-spaced filter set has the advantage that internally measurements can be compared exactly: a supernova will have the same supernova-frame standardized magnitude if it is, for example, observed with the two bluer bands at $z=1.2$ or the two redder bands at $z=1.77$, because the LH1 and LH2 bands cover the same supernova-frame wavelength range at $z=1.2$ as the LH2 and LH3 bands at $z=1.77$. Also, as with the lower-redshift results using logarithmically-spaced filters, a periodic bias found in the high-redshift results using logarithmically-spaced filters can in principle be fit out in a cosmological analysis.

\subsubsection*{Impact on $w$ and the Effect of a Possible Population Shift}
For applications at higher redshifts, a trend in the biases of the K-correction errors that is monotonic with redshift, such as was seen at lower redshifts in the results in using the MegaCam-R09 filters (Section~\ref{sec: Sources}), will be a concern when supernova searches are applied to questions in cosmology, in particular the calculation of the time variation of the dark energy equation of state ratio $w$. Subtracting out the bias found using the MegaCam-R09 filters from data in the range $0<z<0.8$ causes $w$ to shift by about $-0.03$ (using as a representative sample supernovae from the Union Supernova Ia Compilation, Rubin et al., in preparation). Since the bias has been calculated using the supernovae that we artificially redshifted, we could simply correct for it in future searches, but that assumes that the supernova population at higher redshifts is the same as the low redshift population. 

To measure the effects of a change in population, we repeated the calculation of the bias and dispersion on supernova sets where the population shifts towards a certain subset of supernovae as the redshift increases. As a simple test, the supernovae were divided into two sets, one containing supernovae with average standardized magnitude variation (using the magnitudes calculated at redshifts between $z=0.3$ and $0.7$) above the median, called the high-bias set, and one with average standardized magnitude variation below the median, called the low-bias set. We then simulated supernova sets, which at $z=0$ would have the original population of 50\% high-bias supernovae and 50\% low-bias supernovae, and would then shift quadratically with redshift so that at $z=2.0$ they would have 95\% high-bias supernovae and 5\% low-bias supernovae. The bias and dispersion of the supernova sets were calculated with the MegaCam-R09 filters up to $z=0.8$ and with the Euclid-L11 filters from there to $z=2.0$. This was then repeated with supernova sets that shift toward having more low bias supernovae at high redshift. 

\begin{figure}
\includegraphics[width = \columnwidth]{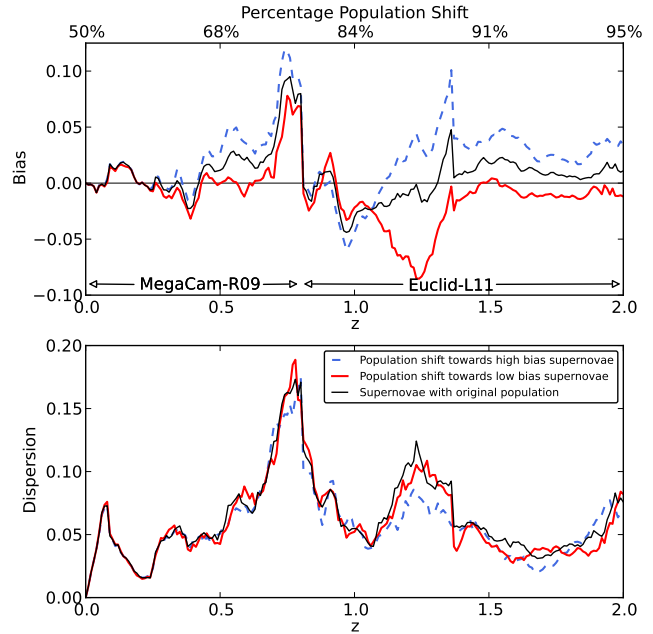}
\caption{Bias and dispersion in the standardized magnitude variation due to K-correction errors in supernova populations shifted towards high-bias supernovae (in blue) or low-bias supernovae (in red) when using SALT2.2. The results from the MegaCam-R09 filters are used from $z=0$ to $z=0.8$ and the results from Euclid-L11 are used from there to $z=2.0$. The population consists of 50\% high-bias and 50\% low-bias supernovae at $z=0$ and shifts quadratically such that in the region of $z=0.6$ to $0.8$ about 70\% of the supernovae are from one population, and at $z=2.0$, 95\% of the supernovae are from one and 5\% are from the other. The results with the original population are shown in black.}
\label{fig: Weighted}
\end{figure}
The results, shown in Figure~\ref{fig: Weighted}, indicate that even at redshifts like $z=0.6$ where the population is only shifted from 50\% to about 70\% towards one population, the difference between the biases is near 0.05 mag. At higher redshifts, the effect is more consistently large, though particularly large at the  redshifts around $z=1.25$ where the filter configuration gives more weight to regions of the spectrum where the two populations differ more. The difference between the bias calculated with the original population and the bias with a population shift would correspond to a change in $w$ of $\sim0.03$ when calculated with data up to $z=0.8$ (again using supernovae from the Union Supernova Ia Compilation). This indicates that without some knowledge of the supernova population, attempts to correct for a bias in magnitudes due to K-correction errors will introduce a significant amount of uncertainty. It is important to note that these biases, too, always become small when there are redshifts at which the low- and high-redshift observer's filters align (e.g. around $z= 1$ in Figure~\ref{fig: Weighted}). While an accurate measurement could be made at these points, this would only be possible for a limited range of redshifts for each filter set, an expensive proposition as a way to study the variation of $w$ with time.

\section{Discussion}
\label{sec: disc}
We have isolated the component of the standardized magnitude variation due to K-corrections across a range of redshifts. These inconsistencies in the magnitudes measured are a direct consequence of the imperfect way in which the lightcurve fitting template is able to match the underlying spectrum of each supernova at each phase. The best fit of an individual supernova varies depending on the filter configuration with which it was observed: examination of the actual spectral time series of a supernova juxtaposed with the spectra constructed from the SALT2 fit of the supernova shows that there are differences between the data and the template fit to it at every redshift. When a fixed filter system is used over a range of redshifts, these differences move across the photometric filters, causing changes in the best fit of the parameters to the lightcurve. As a result we see a variation in the standardized magnitudes fit to the supernova. Biases appear in the average fitted standardized magnitudes when there is an offset between the population of the observed supernovae and the supernovae in the lightcurve fitter's training set. In the current generation of lightcurve fitters, this may be due to the fact that the models are trained on photometry from limited redshift ranges with no spectra or with spectra from a relatively small number of supernovae. The dispersion that is seen in the standardized magnitudes, however, is an inherent effect of the fact that the flexibility of a model built with two linear components and a color law is unable to capture the diversity of Type~Ia supernovae. This is reinforced by the finding in \cite{Kim:2013} that a model incorporating more parameters produced a significant reduction in the dispersion.

We see then that even when using a fairly sophisticated two-parameter lightcurve fitter systematic errors remain. These produce significant, though currently subdominant, amounts of dispersion in supernovae magnitudes, along with bias if the fitter training set does not match the distribution of the fitted supernovae. A more precise model cannot be trained with data insufficient to capture the full range of Type~Ia supernova behaviors. However, the existence of the SNfactory spectrophotometric time series suggests possible methods for improving results. As described in Section~\ref{sec: Apps}, the SNfactory data could be used to predict bias and subtract it out of high-redshift results, but this is limited by our knowledge of the high-redshift population. A more sophisticated approach would be using the SNfactory data as the low-redshift anchor for the Hubble diagram and performing direct matching between the SNfactory supernovae and high redshift supernovae, eliminating the step of using a template-based lightcurve fitter (Fakhouri et al., in preparation). Another possibility would be to use SNfactory data to train a more complex spectral model with the flexibility to capture more of Type Ia supernova diversity, though lightcurves alone may not have sufficient data on which to fit such a model.

This then highlights the primary remaining issues: that we cannot currently recognize which model or observed low-redshift spectrum corresponds to a given high-redshift supernova from photometric light curves alone, and that we cannot assume that the overall population stays the same across redshifts. Future work must determine whether spectroscopy of high-redshift supernovae is necessary or if there are unique photometric indicators beyond the currently employed lightcurve width and color that can be used to identify a given supernova among the diverse range of possible Type~Ia supernovae.

\section{Conclusion}
We have calculated the variation in distance moduli due to K-corrections for filter sets of various configurations and with wavelength coverages selected for low to mid-range and high redshift supernovae. We find that the dispersion and bias in the distance moduli are the most predictable and the most constrained when we use logarithmically-shaped filters that are sufficiently numerous to ensure that the full wavelength range of the supernovae is covered. However, this variation in the distance moduli is a symptom of using a template that lacks the flexibility to span the diversity of Type~Ia supernovae, meaning it will occur at most redshifts regardless of the choice of filters. While we have used the two-parameter lightcurve fitter SALT2 in this analysis, no other current fitters provide different methods that would resolve the issues seen here. Thus, the results show that if we use photometry and simple template-based lightcurve fitters, we must expect that errors in our magnitude measurements will be propagated into cosmology measurements.

A supernova spectral-time-series template trained on low-redshift spectrophotometry from the SNfactory should be capable of making an improvement in the bias of distance moduli due to K-corrections, but would be handicapped by lack of information concerning shifts in the supernova population. In this case then, we would need to spot check our high-redshift photometry for changes in supernova population or environment, which would be expensive. Another possibility for improving results would be redundant, overlapping filters, but this is also expensive. Nevertheless, in order to accurately measure high-redshift supernovae, future searches must have a way to identify and account for population drift, and this may in the end require partial or complete spectrophotometry.

\section*{Acknowledgments}
The authors are grateful to the technical and scientific staff of the University of Hawaii 2.2 m telescope, the Palomar Observatory, and the High Performance Wireless Radio Network (HPWREN). We thank Dan Birchall for his assistance in collecting data with SNIFS. We wish to recognize and acknowledge the very significant cultural role and reverence that the summit of Mauna Kea has always had within the indigenous Hawaiian community. This work was supported by the Director, Office of Science, Office of High Energy Physics, of the U.S. Department of Energy under Contract No. DE-AC02-05CH11231; by a grant from the Gordon \& Betty Moore Foundation; in France by support from CNRS/IN2P3, CNRS/INSU, and PNC; and in Germany by the DFG through TRR33 ``The Dark Universe." NC is grateful to the LABEX Lyon Institute of Origins (ANR-10-LABX-0066) of the Universit\'e de Lyon for its financial support within the program ``Investissements d'Avenir" (ANR-11-IDEX-0007) of the French government operated by the National Research Agency (ANR). National Science Foundation Grant Number ANI-0087344 and the University of California, San Diego provided funding for HPWREN.

\bibliographystyle{apj}
\bibliography{K_corr_bib2}

\begin{thebibliography}{}
\expandafter\ifx\csname natexlab\endcsname\relax\def\natexlab#1{#1}\fi

\bibitem[{Aldering {et~al.}(2002)Aldering, Adam, Antilogus, Astier, Bacon,
  Bongard, Bonnaud, Copin, Hardin, Henault, Howell, Lemonnier, Levy, Nugent,
  Pain, Pecontal, Pecontal, Perlmutter, Quimby, Schamaneche, Smadja, \&
  Wood-Vasey}]{Aldering}
Aldering, G., Adam, G., Antilogus, P., {et~al.} 2002, Society of Photo-Optical
  Instrumentation Engineers (SPIE) Conference Series, 4836, 61

\bibitem[{{Aldering} {et~al.}(2006){Aldering}, {Antilogus}, {Bailey}, {Baltay},
  {Bauer}, {Blanc}, {Bongard}, {Copin}, {Gangler}, {Gilles}, {Kessler},
  {Kocevski}, {Lee}, {Loken}, {Nugent}, {Pain}, {P{\'e}contal}, {Pereira},
  {Perlmutter}, {Rabinowitz}, {Rigaudier}, {Scalzo}, {Smadja}, {Thomas},
  {Wang}, {Weaver}, \& {Nearby Supernova Factory}}]{Aldering:2006}
{Aldering}, G., {Antilogus}, P., {Bailey}, S., {et~al.} 2006, \apj, 650, 510

\bibitem[{{Astier} {et~al.}(2006){Astier}, {Guy}, {Regnault}, {Pain},
  {Aubourg}, {Balam}, {Basa}, {Carlberg}, {Fabbro}, {Fouchez}, {Hook},
  {Howell}, {Lafoux}, {Neill}, {Palanque-Delabrouille}, {Perrett}, {Pritchet},
  {Rich}, {Sullivan}, {Taillet}, {Aldering}, {Antilogus}, {Arsenijevic},
  {Balland}, {Baumont}, {Bronder}, {Courtois}, {Ellis}, {Filiol}, {Gon{\c
  c}alves}, {Goobar}, {Guide}, {Hardin}, {Lusset}, {Lidman}, {McMahon},
  {Mouchet}, {Mourao}, {Perlmutter}, {Ripoche}, {Tao}, \&
  {Walton}}]{Astier:2006}
{Astier}, P., {Guy}, J., {Regnault}, N., {et~al.} 2006, \aap, 447, 31

\bibitem[{{Bacon} {et~al.}(2000){Bacon}, {Emsellem}, {Copin}, \&
  {Monnet}}]{Bacon:2000}
{Bacon}, R., {Emsellem}, E., {Copin}, Y., \& {Monnet}, G. 2000, in Astronomical
  Society of the Pacific Conference Series, Vol. 195, Imaging the Universe in
  Three Dimensions, ed. W.~{van Breugel} \& J.~{Bland-Hawthorn}, 173

\bibitem[{{Bacon} {et~al.}(1995){Bacon}, {Adam}, {Baranne}, {Courtes}, {Dubet},
  {Dubois}, {Emsellem}, {Ferruit}, {Georgelin}, {Monnet}, {Pecontal},
  {Rousset}, \& {Say}}]{Bacon:1995}
{Bacon}, R., {Adam}, G., {Baranne}, A., {et~al.} 1995, \aaps, 113, 347

\bibitem[{{Bacon} {et~al.}(2001){Bacon}, {Copin}, {Monnet}, {Miller},
  {Allington-Smith}, {Bureau}, {Carollo}, {Davies}, {Emsellem}, {Kuntschner},
  {Peletier}, {Verolme}, \& {de Zeeuw}}]{Bacon:2001}
{Bacon}, R., {Copin}, Y., {Monnet}, G., {et~al.} 2001, \mnras, 326, 23

\bibitem[{{Bailey} {et~al.}(2009){Bailey}, {Aldering}, {Antilogus}, {Aragon},
  {Baltay}, {Bongard}, {Buton}, {Childress}, {Chotard}, {Copin}, {Gangler},
  {Loken}, {Nugent}, {Pain}, {Pecontal}, {Pereira}, {Perlmutter}, {Rabinowitz},
  {Rigaudier}, {Runge}, {Scalzo}, {Smadja}, {Swift}, {Tao}, {Thomas}, {Wu}, \&
  {Nearby Supernova Factory}}]{Bailey:2009}
{Bailey}, S., {Aldering}, G., {Antilogus}, P., {et~al.} 2009, \aap, 500, L17

\bibitem[{{Bernstein} {et~al.}(2012){Bernstein}, {Kessler}, {Kuhlmann},
  {Biswas}, {Kovacs}, {Aldering}, {Crane}, {D'Andrea}, {Finley}, {Frieman},
  {Hufford}, {Jarvis}, {Kim}, {Marriner}, {Mukherjee}, {Nichol}, {Nugent},
  {Parkinson}, {Reis}, {Sako}, {Spinka}, \& {Sullivan}}]{Bernstein:2012}
{Bernstein}, J.~P., {Kessler}, R., {Kuhlmann}, S., {et~al.} 2012, \apj, 753,
  152

\bibitem[{{Betoule} {et~al.}(2014){Betoule}, {Kessler}, {Guy}, {Mosher},
  {Hardin}, {Biswas}, {Astier}, {El-Hage}, {Konig}, {Kuhlmann}, {Marriner},
  {Pain}, {Regnault}, {Balland}, {Bassett}, {Brown}, {Campbell}, {Carlberg},
  {Cellier-Holzem}, {Cinabro}, {Conley}, {D'Andrea}, {DePoy}, {Doi}, {Ellis},
  {Fabbro}, {Filippenko}, {Foley}, {Frieman}, {Fouchez}, {Galbany}, {Goobar},
  {Gupta}, {Hill}, {Hlozek}, {Hogan}, {Hook}, {Howell}, {Jha}, {Le Guillou},
  {Leloudas}, {Lidman}, {Marshall}, {M{\"o}ller}, {Mour{\~a}o}, {Neveu},
  {Nichol}, {Olmstead}, {Palanque-Delabrouille}, {Perlmutter}, {Prieto},
  {Pritchet}, {Richmond}, {Riess}, {Ruhlmann-Kleider}, {Sako}, {Schahmaneche},
  {Schneider}, {Smith}, {Sollerman}, {Sullivan}, {Walton}, \&
  {Wheeler}}]{Betoule:2014}
{Betoule}, M., {Kessler}, R., {Guy}, J., {et~al.} 2014, \aap, 568, A22

\bibitem[{{Bohlin}(2014)}]{Bohlin:2014}
{Bohlin}, R.~C. 2014, \aj, 147, 127

\bibitem[{{Bongard} {et~al.}(2011){Bongard}, {Soulez}, {Thi{\'e}baut}, \&
  {Pecontal}}]{Bongard:2011}
{Bongard}, S., {Soulez}, F., {Thi{\'e}baut}, {\'E}., \& {Pecontal}, {\'E}.
  2011, \mnras, 418, 258

\bibitem[{Burns {et~al.}(2010)Burns, Stritzinger, Phillips, Kattner, Persson,
  Madore, Freedman, Boldt, Campillay, Contreras, Folatelli, Gonzalez,
  Krzeminski, Morrell, Salgado, \& Suntzeff}]{Burns:2010vn}
Burns, C.~R., Stritzinger, M., Phillips, M.~M., {et~al.} 2010, Astronomical
  Journal, 1010.4040

\bibitem[{{Buton} {et~al.}(2013){Buton}, {Copin}, {Aldering}, {Antilogus},
  {Aragon}, {Bailey}, {Baltay}, {Bongard}, {Canto}, {Cellier-Holzem},
  {Childress}, {Chotard}, {Fakhouri}, {Gangler}, {Guy}, {Hsiao}, {Kerschhaggl},
  {Kowalski}, {Loken}, {Nugent}, {Paech}, {Pain}, {P{\'e}contal}, {Pereira},
  {Perlmutter}, {Rabinowitz}, {Rigault}, {Runge}, {Scalzo}, {Smadja}, {Tao},
  {Thomas}, {Weaver}, {Wu}, \& {Nearby SuperNova Factory}}]{Buton:2012}
{Buton}, C., {Copin}, Y., {Aldering}, G., {et~al.} 2013, \aap, 549, A8

\bibitem[{{Cardelli} {et~al.}(1989){Cardelli}, {Clayton}, \&
  {Mathis}}]{Cardelli:1989}
{Cardelli}, J.~A., {Clayton}, G.~C., \& {Mathis}, J.~S. 1989, \apj, 345, 245

\bibitem[{{Childress} {et~al.}(2013){Childress}, {Aldering}, {Antilogus},
  {Aragon}, {Bailey}, {Baltay}, {Bongard}, {Buton}, {Canto}, {Cellier-Holzem},
  {Chotard}, {Copin}, {Fakhouri}, {Gangler}, {Guy}, {Hsiao}, {Kerschhaggl},
  {Kim}, {Kowalski}, {Loken}, {Nugent}, {Paech}, {Pain}, {Pecontal}, {Pereira},
  {Perlmutter}, {Rabinowitz}, {Rigault}, {Runge}, {Scalzo}, {Smadja}, {Tao},
  {Thomas}, {Weaver}, \& {Wu}}]{Childress:2013}
{Childress}, M., {Aldering}, G., {Antilogus}, P., {et~al.} 2013, \apj, 770, 107

\bibitem[{{Chotard} {et~al.}(2011){Chotard}, {Gangler}, {Aldering},
  {Antilogus}, {Aragon}, {Bailey}, {Baltay}, {Bongard}, {Buton}, {Canto},
  {Childress}, {Copin}, {Fakhouri}, {Hsiao}, {Kerschhaggl}, {Kowalski},
  {Loken}, {Nugent}, {Paech}, {Pain}, {Pecontal}, {Pereira}, {Perlmutter},
  {Rabinowitz}, {Runge}, {Scalzo}, {Smadja}, {Tao}, {Thomas}, {Weaver}, {Wu},
  \& {Nearby Supernova Factory}}]{Chotard:2011}
{Chotard}, N., {Gangler}, E., {Aldering}, G., {et~al.} 2011, \aap, 529, L4

\bibitem[{Conley {et~al.}(2008)Conley, Sullivan, Hsiao, Guy, Astier, Balam,
  Balland, Basa, Carlberg, Fouchez, Hardin, Howell, Hook, Pain, Perrett,
  Pritchet, \& Regnault}]{Conley:2008kx}
Conley, A., Sullivan, M., Hsiao, E.~Y., {et~al.} 2008, Astrophysics Journal,
  0803.3441

\bibitem[{{Ellis} {et~al.}(2008){Ellis}, {Sullivan}, {Nugent}, {Howell},
  {Gal-Yam}, {Astier}, {Balam}, {Balland}, {Basa}, {Carlberg}, {Conley},
  {Fouchez}, {Guy}, {Hardin}, {Hook}, {Pain}, {Perrett}, {Pritchet}, \&
  {Regnault}}]{Ellis:2008}
{Ellis}, R.~S., {Sullivan}, M., {Nugent}, P.~E., {et~al.} 2008, \apj, 674, 51

\bibitem[{{Frieman} {et~al.}(2008){Frieman}, {Bassett}, {Becker}, {Choi},
  {Cinabro}, {DeJongh}, {Depoy}, {Dilday}, {Doi}, {Garnavich}, {Hogan},
  {Holtzman}, {Im}, {Jha}, {Kessler}, {Konishi}, {Lampeitl}, {Marriner},
  {Marshall}, {McGinnis}, {Miknaitis}, {Nichol}, {Prieto}, {Riess}, {Richmond},
  {Romani}, {Sako}, {Schneider}, {Smith}, {Takanashi}, {Tokita}, {van der
  Heyden}, {Yasuda}, {Zheng}, {Adelman-McCarthy}, {Annis}, {Assef},
  {Barentine}, {Bender}, {Blandford}, {Boroski}, {Bremer}, {Brewington},
  {Collins}, {Crotts}, {Dembicky}, {Eastman}, {Edge}, {Edmondson}, {Elson},
  {Eyler}, {Filippenko}, {Foley}, {Frank}, {Goobar}, {Gueth}, {Gunn},
  {Harvanek}, {Hopp}, {Ihara}, {Ivezi{\'c}}, {Kahn}, {Kaplan}, {Kent},
  {Ketzeback}, {Kleinman}, {Kollatschny}, {Kron}, {Krzesi{\'n}ski}, {Lamenti},
  {Leloudas}, {Lin}, {Long}, {Lucey}, {Lupton}, {Malanushenko}, {Malanushenko},
  {McMillan}, {Mendez}, {Morgan}, {Morokuma}, {Nitta}, {Ostman}, {Pan},
  {Rockosi}, {Romer}, {Ruiz-Lapuente}, {Saurage}, {Schlesinger}, {Snedden},
  {Sollerman}, {Stoughton}, {Stritzinger}, {Subba Rao}, {Tucker}, {Vaisanen},
  {Watson}, {Watters}, {Wheeler}, {Yanny}, \& {York}}]{Frieman}
{Frieman}, J.~A., {Bassett}, B., {Becker}, A., {et~al.} 2008, \aj, 135, 338

\bibitem[{Guy {et~al.}(2007)Guy, Astier, Baumont, Hardin, Pain, Regnault, Basa,
  Carlberg, Conley, Fabbro, Fouchez, Hook, Howell, Perrett, Pritchet, Rich,
  Sullivan, Antilogus, Aubourg, Bazin, Bronder, Filiol, Palanque-Delabrouille,
  Ripoche, \& Ruhlmann-Kleider}]{Guy:2007fk}
Guy, J., Astier, P., Baumont, S., {et~al.} 2007, Astron.Astrophys., 466, 11

\bibitem[{{Guy} {et~al.}(2010){Guy}, {Sullivan}, {Conley}, {Regnault},
  {Astier}, {Balland}, {Basa}, {Carlberg}, {Fouchez}, {Hardin}, {Hook},
  {Howell}, {Pain}, {Palanque-Delabrouille}, {Perrett}, {Pritchet}, {Rich},
  {Ruhlmann-Kleider}, {Balam}, {Baumont}, {Ellis}, {Fabbro}, {Fakhouri},
  {Fourmanoit}, {Gonz{\'a}lez-Gait{\'a}n}, {Graham}, {Hsiao}, {Kronborg},
  {Lidman}, {Mourao}, {Perlmutter}, {Ripoche}, {Suzuki}, \&
  {Walker}}]{Guy:2010}
{Guy}, J., {Sullivan}, M., {Conley}, A., {et~al.} 2010, \aap, 523, A7

\bibitem[{{Hamuy} {et~al.}(1994){Hamuy}, {Suntzeff}, {Heathcote}, {Walker},
  {Gigoux}, \& {Phillips}}]{Hamuy:1994}
{Hamuy}, M., {Suntzeff}, N.~B., {Heathcote}, S.~R., {et~al.} 1994, \pasp, 106,
  566

\bibitem[{{Hamuy} {et~al.}(1992){Hamuy}, {Walker}, {Suntzeff}, {Gigoux},
  {Heathcote}, \& {Phillips}}]{Hamuy:1992}
{Hamuy}, M., {Walker}, A.~R., {Suntzeff}, N.~B., {et~al.} 1992, \pasp, 104, 533

\bibitem[{{Holtzman} {et~al.}(2008){Holtzman}, {Marriner}, {Kessler}, {Sako},
  {Dilday}, {Frieman}, {Schneider}, {Bassett}, {Becker}, {Cinabro}, {DeJongh},
  {Depoy}, {Doi}, {Garnavich}, {Hogan}, {Jha}, {Konishi}, {Lampeitl},
  {Marshall}, {McGinnis}, {Miknaitis}, {Nichol}, {Prieto}, {Riess}, {Richmond},
  {Romani}, {Smith}, {Takanashi}, {Tokita}, {van der Heyden}, {Yasuda}, \&
  {Zheng}}]{Holtzman}
{Holtzman}, J.~A., {Marriner}, J., {Kessler}, R., {et~al.} 2008, \aj, 136, 2306

\bibitem[{{Humason} {et~al.}(1956){Humason}, {Mayall}, \&
  {Sandage}}]{Humason:1956}
{Humason}, M.~L., {Mayall}, N.~U., \& {Sandage}, A.~R. 1956, \aj, 61, 97

\bibitem[{Jha {et~al.}(2007)Jha, Riess, \& Kirshner}]{Jha:2007ys}
Jha, S., Riess, A.~G., \& Kirshner, R.~P. 2007, Astrophys.J., 659, 122

\bibitem[{{Jha} {et~al.}(2007){Jha}, {Riess}, \& {Kirshner}}]{Jha:2007}
{Jha}, S., {Riess}, A.~G., \& {Kirshner}, R.~P. 2007, \apj, 659, 122

\bibitem[{{Kim} {et~al.}(1996){Kim}, {Goobar}, \& {Perlmutter}}]{Kim:1996}
{Kim}, A., {Goobar}, A., \& {Perlmutter}, S. 1996, \pasp, 108, 190

\bibitem[{{Kim} {et~al.}(2013){Kim}, {Thomas}, {Aldering}, {Antilogus},
  {Aragon}, {Bailey}, {Baltay}, {Bongard}, {Buton}, {Canto}, {Cellier-Holzem},
  {Childress}, {Chotard}, {Copin}, {Fakhouri}, {Gangler}, {Guy}, {Kerschhaggl},
  {Kowalski}, {Nordin}, {Nugent}, {Paech}, {Pain}, {Pecontal}, {Pereira},
  {Perlmutter}, {Rabinowitz}, {Rigault}, {Runge}, {Saunders}, {Scalzo},
  {Smadja}, {Tao}, {Weaver}, \& {Wu}}]{Kim:2013}
{Kim}, A.~G., {Thomas}, R.~C., {Aldering}, G., {et~al.} 2013, \apj, 766, 84

\bibitem[{{Lantz} {et~al.}(2004){Lantz}, {Aldering}, {Antilogus}, {Bonnaud},
  {Capoani}, {Castera}, {Copin}, {Dubet}, {Gangler}, {Henault}, {Lemonnier},
  {Pain}, {Pecontal}, {Pecontal}, \& {Smadja}}]{Lantz:2004}
{Lantz}, B., {Aldering}, G., {Antilogus}, P., {et~al.} 2004, in Society of
  Photo-Optical Instrumentation Engineers (SPIE) Conference Series, Vol. 5249,
  Society of Photo-Optical Instrumentation Engineers (SPIE) Conference Series,
  ed. L.~{Mazuray}, P.~J. {Rogers}, \& R.~{Wartmann}, 146--155

\bibitem[{{Laureijs} {et~al.}(2011){Laureijs}, {Amiaux}, {Arduini},
  {Augu{\`e}res}, {Brinchmann}, {Cole}, {Cropper}, {Dabin}, {Duvet}, {Ealet},
  \& et~al.}]{Laureijs:2011}
{Laureijs}, R., {Amiaux}, J., {Arduini}, S., {et~al.} 2011, ArXiv e-prints,
  arXiv:1110.3193

\bibitem[{{LSST Science Collaboration} {et~al.}(2009){LSST Science
  Collaboration}, {Abell}, {Allison}, {Anderson}, {Andrew}, {Angel}, {Armus},
  {Arnett}, {Asztalos}, {Axelrod}, \& et~al.}]{LSST:2009}
{LSST Science Collaboration}, {Abell}, P.~A., {Allison}, J., {et~al.} 2009,
  ArXiv e-prints, arXiv:0912.0201

\bibitem[{{Maguire} {et~al.}(2012){Maguire}, {Sullivan}, {Ellis}, {Nugent},
  {Howell}, {Gal-Yam}, {Cooke}, {Mazzali}, {Pan}, {Dilday}, {Thomas}, {Arcavi},
  {Ben-Ami}, {Bersier}, {Bianco}, {Fulton}, {Hook}, {Horesh}, {Hsiao}, {James},
  {Podsiadlowski}, {Walker}, {Yaron}, {Kasliwal}, {Laher}, {Law}, {Ofek},
  {Poznanski}, \& {Surace}}]{Maguire:2012}
{Maguire}, K., {Sullivan}, M., {Ellis}, R.~S., {et~al.} 2012, \mnras, 426, 2359

\bibitem[{{Mosher} {et~al.}(2014){Mosher}, {Guy}, {Kessler}, {Astier},
  {Marriner}, {Betoule}, {Sako}, {El-Hage}, {Biswas}, {Pain}, {Kuhlmann},
  {Regnault}, {Frieman}, \& {Schneider}}]{Mosher:2014}
{Mosher}, J., {Guy}, J., {Kessler}, R., {et~al.} 2014, ArXiv e-prints,
  arXiv:1401.4065

\bibitem[{Nugent {et~al.}(2002)Nugent, Kim, \& Perlmutter}]{Nugent:2002fk}
Nugent, P., Kim, A., \& Perlmutter, S. 2002, Publ.Astron.Soc.Pac., 114, 803

\bibitem[{{Oke} \& {Sandage}(1968)}]{Oke:1968}
{Oke}, J.~B., \& {Sandage}, A. 1968, \apj, 154, 21

\bibitem[{{Regnault} {et~al.}(2009){Regnault}, {Conley}, {Guy}, {Sullivan},
  {Cuillandre}, {Astier}, {Balland}, {Basa}, {Carlberg}, {Fouchez}, {Hardin},
  {Hook}, {Howell}, {Pain}, {Perrett}, \& {Pritchet}}]{Regnault:2009}
{Regnault}, N., {Conley}, A., {Guy}, J., {et~al.} 2009, \aap, 506, 999

\bibitem[{{Rubin} {et~al.}(2013){Rubin}, {Knop}, {Rykoff}, {Aldering},
  {Amanullah}, {Barbary}, {Burns}, {Conley}, {Connolly}, {Deustua}, {Fadeyev},
  {Fakhouri}, {Fruchter}, {Gibbons}, {Goldhaber}, {Goobar}, {Hsiao}, {Huang},
  {Kowalski}, {Lidman}, {Meyers}, {Nordin}, {Perlmutter}, {Saunders},
  {Spadafora}, {Stanishev}, {Suzuki}, {Wang}, \& {Supernova Cosmology
  Project}}]{Rubin:2013}
{Rubin}, D., {Knop}, R.~A., {Rykoff}, E., {et~al.} 2013, \apj, 763, 35

\bibitem[{{Scalzo} {et~al.}(2012){Scalzo}, {Aldering}, {Antilogus}, {Aragon},
  {Bailey}, {Baltay}, {Bongard}, {Buton}, {Canto}, {Cellier-Holzem},
  {Childress}, {Chotard}, {Copin}, {Fakhouri}, {Gangler}, {Guy}, {Hsiao},
  {Kerschhaggl}, {Kowalski}, {Nugent}, {Paech}, {Pain}, {Pecontal}, {Pereira},
  {Perlmutter}, {Rabinowitz}, {Rigault}, {Runge}, {Smadja}, {Tao}, {Thomas},
  {Weaver}, {Wu}, \& {Nearby Supernova Factory}}]{Scalzo:2012}
{Scalzo}, R., {Aldering}, G., {Antilogus}, P., {et~al.} 2012, \apj, 757, 12

\bibitem[{{Scalzo} {et~al.}(2010){Scalzo}, {Aldering}, {Antilogus}, {Aragon},
  {Bailey}, {Baltay}, {Bongard}, {Buton}, {Childress}, {Chotard}, {Copin},
  {Fakhouri}, {Gal-Yam}, {Gangler}, {Hoyer}, {Kasliwal}, {Loken}, {Nugent},
  {Pain}, {P{\'e}contal}, {Pereira}, {Perlmutter}, {Rabinowitz}, {Rau},
  {Rigaudier}, {Runge}, {Smadja}, {Tao}, {Thomas}, {Weaver}, \&
  {Wu}}]{Scalzo:2010}
{Scalzo}, R.~A., {Aldering}, G., {Antilogus}, P., {et~al.} 2010, \apj, 713,
  1073

\bibitem[{{Schlegel} {et~al.}(1998){Schlegel}, {Finkbeiner}, \&
  {Davis}}]{Schlegel:1998}
{Schlegel}, D.~J., {Finkbeiner}, D.~P., \& {Davis}, M. 1998, \apj, 500, 525

\bibitem[{{Schmidt} {et~al.}(1998){Schmidt}, {Suntzeff}, {Phillips},
  {Schommer}, {Clocchiatti}, {Kirshner}, {Garnavich}, {Challis}, {Leibundgut},
  {Spyromilio}, {Riess}, {Filippenko}, {Hamuy}, {Smith}, {Hogan}, {Stubbs},
  {Diercks}, {Reiss}, {Gilliland}, {Tonry}, {Maza}, {Dressler}, {Walsh}, \&
  {Ciardullo}}]{Schmidt:1998}
{Schmidt}, B.~P., {Suntzeff}, N.~B., {Phillips}, M.~M., {et~al.} 1998, \apj,
  507, 46

\bibitem[{{Thomas} {et~al.}(2007){Thomas}, {Aldering}, {Antilogus}, {Aragon},
  {Bailey}, {Baltay}, {Baron}, {Bauer}, {Buton}, {Bongard}, {Copin}, {Gangler},
  {Gilles}, {Kessler}, {Loken}, {Nugent}, {Pain}, {Parrent}, {P{\'e}contal},
  {Pereira}, {Perlmutter}, {Rabinowitz}, {Rigaudier}, {Runge}, {Scalzo},
  {Smadja}, {Wang}, {Weaver}, \& {Nearby Supernova Factory}}]{Thomas:2007}
{Thomas}, R.~C., {Aldering}, G., {Antilogus}, P., {et~al.} 2007, \apjl, 654,
  L53

\bibitem[{{Thomas} {et~al.}(2011){Thomas}, {Aldering}, {Antilogus}, {Aragon},
  {Bailey}, {Baltay}, {Bongard}, {Buton}, {Canto}, {Childress}, {Chotard},
  {Copin}, {Fakhouri}, {Gangler}, {Hsiao}, {Kerschhaggl}, {Kowalski}, {Loken},
  {Nugent}, {Paech}, {Pain}, {Pecontal}, {Pereira}, {Perlmutter}, {Rabinowitz},
  {Rigault}, {Rubin}, {Runge}, {Scalzo}, {Smadja}, {Tao}, {Weaver}, {Wu},
  {Brown}, {Milne}, \& {Nearby Supernova Factory}}]{Thomas:2011}
---. 2011, \apj, 743, 27

\end{thebibliography}

\end{document}